\documentclass[a4paper]{article}

\usepackage[T1]{fontenc}
\usepackage[utf8]{inputenc}
\usepackage[english]{babel}
\usepackage{graphics}
\usepackage{epsfig}
\usepackage{amsmath}
\usepackage{amssymb}
\usepackage{amsthm}
\usepackage{url}
\usepackage[breaklinks]{hyperref}
\usepackage{float}
\usepackage{pifont}
\usepackage[table]{xcolor}
\usepackage{lmodern}
\usepackage{xspace}
\usepackage{bbold}
\usepackage[capitalize,nameinlink]{cleveref}

\usepackage{draftwatermark} 
\SetWatermarkText{Preprint} 
\SetWatermarkScale{1}       

\usepackage{graphicx}
\usepackage{mathtools}
\usepackage[english]{babel}
\usepackage{amsmath,amsfonts,amssymb}

\usepackage{hyperref}

\usepackage{subcaption}

\newcommand{\R}{\mathbb{R}}
\DeclareMathOperator*{\argmin}{arg\,min}

\newcommand{\frosym}{\mathrm{F}}
\newcommand{\fro}{_\frosym}

\begin{document}

\title{\mbox{\hskip-4ex On Reduced Input-Output Dynamic Mode Decomposition}}



\author{Peter~Benner
\thanks{Computational Methods in Systems and Control Theory, Max Planck Institute for Dynamics of Complex Technical Systems, Sandtorstr. 1, 39106 Magdeburg, Germany \newline ORCID: 0000-0003-3362-4103, \url{benner@mpi-magdeburg.mpg.de}} \and
        Christian~Himpe\thanks{Computational Methods in Systems and Control Theory, Max Planck Institute for Dynamics of Complex Technical Systems, Sandtorstr. 1, 39106 Magdeburg, Germany \newline ORCID: 0000-0003-2194-6754, \url{himpe@mpi-magdeburg.mpg.de}} \and
        Tim~Mitchell \thanks{Computational Methods in Systems and Control Theory, Max Planck Institute for Dynamics of Complex Technical Systems, Sandtorstr. 1, 39106 Magdeburg, Germany \newline ORCID: 0000-0002-8426-0242, \url{mitchell@mpi-magdeburg.mpg.de}}
}


\date{}

\maketitle

\begin{abstract}\setlength{\parindent}{0pt}
The identification of reduced-order models from high-dimensional data is a challenging task,
and even more so if the identified system should not only be suitable for a certain data set,
but generally approximate the input-output behavior of the data source.
In this work, we consider the input-output dynamic mode decomposition method for system identification.
We compare excitation approaches for the data-driven identification process
and describe an optimization-based stabilization strategy for the identified systems.

~\\

\textbf{Keywords:} Dynamic Mode Decomposition, Model Reduction, \\ \phantom{\textbf{Keywords: }}System Identification, Cross Gramian, Optimization

~\\

\textbf{MSC:} 93B30, 90C99
\end{abstract}

\vfill

\pagebreak[4]


\section{Introduction}\label{sec:intro}
In various applications, it can be difficult (sometimes even impossible) to derive models from first principles. 
However, the input-output response of a system and data of the inner state may still be available; 
we refer to such a setup as a \emph{graybox}.
For example, a natural process whose underlying action is not well understood can be considered a graybox since we
may only be able to measure its behavior.
In applications, such as manufacturing and design, it may be necessary to model a third-party provided subcomponent 
whose exact or full specifications may not be obtainable due to it containing proprietary information.
In order to gain insight into such natural or technical processes, and derive models that simulate and/or predict their behaviors, 
it is often beneficial and perhaps necessary to create models using measured or generated data. 
The discipline of system identification investigates methods for the task of obtaining such models from data.
One class of methods for data-driven identification is dynamic mode decomposition (DMD), which also provides a modal analysis of the resulting systems. 
In this work, we investigate variants of DMD for the class of input-output systems and compare data sampling strategies.

DMD has its roots in the modal decomposition of the Koopman operator \cite{Koo31,morRowMBetal09},
which has been recently rediscovered for the spectral analysis of fluid dynamics \cite{morMez05}.
The basic DMD method was introduced in \cite{morSch10}, 
and various extensions have been added, such as Exact DMD \cite{morTuRLetal14} or Extended DMD \cite{morCheTR12}.
Furthermore, DMD can also be used as a tool for model order reduction: \cite{morRowD17} proposed using DMD for flow analysis and control while DMD has also been combined with Galerkin-projection techniques to model nonlinear systems \cite{morAllK17}.
For a comprehensive survey of DMD and its variants, see \cite{morKutBBetal16}.

Our work here builds upon two specific variants of DMD.
The first is Dynamic Mode Decomposition with Control (DMDc) \cite{morProBK16}, and thus by relation, also Koopman with Inputs and Control (KIC) \cite{morProBK16a}. 
The second is Input-Output Dynamic Mode Decomposition (ioDMD) \cite{morAnnGS16,morAnnS17}, which itself is closely related to the Direct Numerical Algorithm for Sub-Space State System Identification (N4SID) \cite{Vib95}.
DMDc extends DMD to scenarios where the input of the discrete system approximation is given by a functional while ioDMD additionally handles the case when the system's output is specified and also a functional.

To generically identify a system without prior knowledge on the relevant input functions,
techniques such as persistent excitation \cite{AstE71} have been well known for some time now.
We propose an extension to the ioDMD method of \cite{morAnnGS16} 
with a new excitation variant related to the cross Gramian matrix \cite{morFerN83}.
Additionally, since DMD-based identification does not guarantee that its resulting models are stable,
we propose a post-processing procedure to stabilize ioDMD-derived models, by employing the nonsmooth constrained optimization method of \cite{CurMO17} to solve a corresponding stabilization problem.

This document is structured as follows.
In \cref{sec:dmd}, an overview of the prerequisite dynamic mode decomposition method and its relevant variants is given.
\cref{sec:dat} describes the new excitation strategy while our optimization-based stabilization procedure is discussed in \cref{sec:stab}.
Finally, two numerical experiments are conducted in \cref{sec:numex}.



\section{Dynamic Mode Decomposition}\label{sec:dmd}
Consider the time-invariant ordinary differential equation (ODE):
\begin{align}\label{eq:contsys1}
 \dot{x}(t) = f(x(t)),
\end{align}
with state $x(t) \in \R^N$ and vector field $f:\R^N \to \R^N$.
Furthermore, consider for now that \eqref{eq:contsys1} is sampled at uniform intervals for times $t_0,\ldots,t_K$.
The most basic version of DMD \cite{morSch10,morKutBBetal16} aims to approximate \eqref{eq:contsys1} by constructing a discrete-time linear system
\begin{align}\label{eq:discsys1}
 x_{k+1} = A x_k,
\end{align}
with a linear operator $A \in \R^{N \times N}$, such that if $x_k = x(t_k)$, then $x_{k+1} \approx x(t_{k+1})$ should also hold, for all $k = 0,\ldots,K-1$.

Starting at an initial state $x_0$, the sequence defined by \eqref{eq:discsys1} corresponds to a trajectory of the state vectors $x_k$.
The corresponding data matrix $X  \in \R^{N \times K}$ is simply the in-order concatenation of these state vectors, that is,
\[
	X = \begin{bmatrix} x_0 & x_1 & \dots & x_K \end{bmatrix}.
\]
By forming the following two partial trajectories:
\begin{align*}
 X_0 &= \begin{bmatrix} x_0 & x_1 & \dots & x_{K-1} \end{bmatrix} \quad \text{and} \\
 X_1 &= \begin{bmatrix} x_1 & x_2 & \dots & x_K \end{bmatrix},
\end{align*}
where $X_0 \in \R^{N \times K-1}$ is just the data matrix $X$ with its last data point removed while $X_1 \in \R^{N \times K-1}$ is just $X$ with its first data point removed,
the idea behind (plain) DMD \cite{morSch10} is to then solve the linear system of equations:
\begin{align*}
 X_1 &= A X_0,
\end{align*}
which is in fact just \eqref{eq:discsys1}, where $x_{k+1}$ and $x_k$ have been replaced by $X_1$ and $X_0$, respectively.
A best-fit solution to this problem is given by the pseudoinverse: \begin{align*}
 A \approx X_1 X_0^+,
\end{align*}
which is also the solution to minimizing the least-squares error in the Frobenius norm ($\| \cdot \|\fro$):
\begin{align*}
 A = \argmin_{\tilde{A} \in \R^{N \times N}}(\|X_1 - \tilde{A} X_0\|\fro).
\end{align*}
The DMD modes of \eqref{eq:contsys1} are given by the eigenvectors of the matrix $A$:
\begin{align*}
 A\Lambda = \Lambda V.
\end{align*}

Beyond just using a single uniformly-sampled time series, DMD has also been generalized to a method called Exact DMD \cite{morTuRLetal14},
which can additionally support the concatenation of multiple different time series and/or non-uniform time steppings.
This generalization of DMD is made possible by reducing the requirements of $X_0$ and $X_1$ to the lesser condition that their columns need only be composed in pairs of data such that \mbox{$X_{1}(:,k) = f(X_{0}(:,k))$} is fulfilled.


\subsection{Practical Computation}\label{sec:alg1}
We now give a high-level algorithmic description of DMD identification.
The pseudoinverse of the data matrix can be computed using the (rank-revealing) singular value decomposition (SVD),
$X_0 = \mathcal{U} \Sigma \mathcal{V}^*$, as follows:
\begin{align*}
X_0^+ = \mathcal{V} \Sigma^{-1} \mathcal{U}^*.
\end{align*}
However, inverting tiny but nonzero singular values in the \emph{computed} diagonal matrix $\Sigma$ poses a numerical problem, 
as these small singular values may be inaccurate.  
Applying $\Sigma^{-1}$ could \emph{overamplify} components of $X_0$, in particular, the less important ones.
To counteract this effect, computing the pseudoinverse via the SVD is done in practice by truncating any singular values that are smaller than some fixed $\varepsilon \in \R^+$,
which is equivalent to adding a regularization term to the least-squares solution for $A$:
\begin{align*}
 A = \argmin_{\tilde{A} \in \R^{N \times N}} \|X_1 - \tilde{A} X_0\|\fro^2 + \beta \|\tilde A\|\fro^2,
\end{align*}
for some value of the regularization parameter $\beta \in \R^+$.
Following \cite{morTuRLetal14}, the algorithm for the DMD-based computation of the system matrix $A$, given a state-space trajectory $X$ and a lower bound $\varepsilon > 0$ for discarding tiny singular values, is as follows:
\begin{enumerate}
 \item Sample the state-space trajectory and form data matrix $X = \begin{bmatrix} x_0 & \dots & x_K \end{bmatrix}$.
 \item Assemble the shifted partitions $X_0$ and $X_1$.
 \item Compute the SVD of $X_0 = \mathcal{U} \Sigma \mathcal{V}^*$.
 \item Truncate $\Sigma$: $\widetilde\Sigma_{ii} = \Sigma_{ii}$ if $\Sigma_{ii} \ge \varepsilon$; 0 otherwise.
 \item Identify the approximate discrete system matrix: $A := \mathcal{U}^* X_1 \mathcal{V} \widetilde{\Sigma}^{-1}$
\end{enumerate}
In this DMD variant, the order (dimension) of the computed matrix $A$ is equal to the number of retained (nontruncated) singular values,
but this truncation is done solely for numerical accuracy; 
the intention is to keep as many components of the dynamics as possible.
In contrast, model order reduction typically aims to \emph{significantly} reduce the system down to just a small set of \emph{essential} dynamics,
and accomplishing this goal will be the focus of \cref{sec:rom}.


\subsection{Dynamic Mode Decomposition with Control}\label{sec:dmdc}
Considering systems whose vector field $f$ depends not just on the state but also on an input function $u : \R \to \R^M$:
\begin{align}\label{eq:contsys2}
 \dot{x}(t) = f(x(t),u(t)),
\end{align}
has led to a DMD variant called Dynamic Mode Decomposition with Control (DMDc) \cite{morProBK16}.  
We focus on a specific DMDc variant \cite[Sec.~3.3]{morProBK16} that aims to approximate \eqref{eq:contsys2} by a linear discrete-time control system:
\begin{align}\label{eq:discsys2}
 x_{k+1} = Ax_k + Bu_k,
\end{align}
where $B \in \R^{N \times M}$ (called the input operator) must also be identified in addition to $A$ and input $u_k = u(t_k)$ is a 
discretely-sampled version of the continuous input function $u(t)$ for some sampling times given by $t_k$.
In addition to forming $X$ and $X_0$ as in plain DMD (\cref{sec:dmd}), an analogous construction of matrices for input data is also done.  
An in-order concatenation of the series of input data $u_k$ vectors is done to obtain matrix $U \in \R^{M \times K}$ 
while the partial data matrix $U_0 \in \R^{M  \times K-1}$ is simply $U$ without its last column (the last input sample):
\begin{align*}
 U &= \begin{bmatrix} u_0 & u_1 & \dots & u_{K-1} & u_k \end{bmatrix} \quad \text{and}\\
 U_0 &= \begin{bmatrix} u_0 & u_1 & \dots & u_{K-1} \end{bmatrix}.
\end{align*}
Then, $A$ and $B$ can be obtained by computing the approximate solutions to the linear system of equations given by:
\begin{align*}
 X_1 &= AX_0 + BU_0 = \begin{bmatrix} A & B \end{bmatrix} \begin{bmatrix} X_0 \\ U_0 \end{bmatrix},
\end{align*}
which is \eqref{eq:discsys2} with $u_k$ replaced by $U$, $x_k$ by $X_0$, and $x_{k+1}$ by $X_1$,
and has solution:
\[
 	\begin{bmatrix}A & B \end{bmatrix} = X_1 \begin{bmatrix} X_0 \\ U_0 \end{bmatrix}^+.
\]

As mentioned in \cref{sec:intro}, DMDc is actually a special case of the KIC method \cite{morProBK16a}.
For KIC, the state of the system is also augmented with the discretized input $u_k$, 
which leads the resulting augmented system to have additional operators:
\begin{align*}
 \begin{bmatrix} x_{k+1} \\ u_{k+1} \end{bmatrix} = \begin{bmatrix} A & B \\ B_u & A_u \end{bmatrix} \begin{bmatrix} x_k \\ u_k \end{bmatrix},
\end{align*}
where $B_u \in \R^{M \times N}$ and $A_u \in \R^{M \times M}$.  
Of course, these two additional operators must now also be identified along with matrices $A$ and $B$. 
However, if one assumes that the input is known and no identification of the associated operators is required, then $B_u$ and $A_u$ are just zero matrices, and it is thus clear that KIC is a generalization of DMDc.


\subsection{Input-Output Dynamic Mode Decomposition}
Extending the underlying system once more to now also include an output function $y: \R \to \R^Q$, with an associated output functional $g:\R^N \times \R^M \to \R^Q$ that also depends on the state $x$ and the input $u$, yields the following system:
\begin{align}\label{eq:contsys3}
\begin{split}
 \dot{x}(t) &= f(x(t),u(t)), \\
       y(t) &= g(x(t),u(t)).
\end{split}
\end{align}
Data-based modeling of systems of the form given by \eqref{eq:contsys3} has given rise to a class of DMD methods called Input-Output Dynamic Mode Decomposition (ioDMD) \cite{morAnnGS16}.
Similar to previously discussed DMD variants, the ioDMD method also approximates the original system by a discrete-time linear system, but now with input and output:
\begin{align}\label{eq:discsys3}
\begin{split}
 x_{k+1} &= Ax_k + Bu_k, \\
 y_{k+1} &= Cx_k + Du_k,
\end{split}
\end{align}
where $C \in \R^{Q \times N}$ and $D \in \R^{Q \times M}$ are the output and feed-through operators, respectively.
Since this approximation includes output data,
the discrete output instances $y_k = y(t_k) \in \R^Q$ are also correspondingly arranged into a matrix $Y \in \R^{Q \times K}$ by in-order concatenation while $Y_0 \in \R^{Q \times K-1}$ simply omits the last column (output sample) of $Y$:
\begin{align*}
 Y &= \begin{bmatrix} y_0 & y_1 & \dots & y_{K-1} & y_K \end{bmatrix} \quad \text{and}\\
 Y_0 &= \begin{bmatrix} y_0 & y_1 & \dots & y_{K-1} \end{bmatrix}.
\end{align*}
Matrices $A$, $B$, $C$, and $D$ are then approximated by solving:
\begin{equation}
\label{eq:iodmd_eqns}
 \begin{cases} X_1 = AX_0 + BU_0 \\ Y_0 = CX_0 + DU_0 \end{cases} \quad \text{or equivalently} \quad
 \begin{bmatrix} X_1 \\ Y_0 \end{bmatrix} = \begin{bmatrix} A & B \\ C & D \end{bmatrix} \begin{bmatrix} X_0 \\ U_0 \end{bmatrix},
\end{equation}
which is \eqref{eq:discsys3} but where $u_k$, $x_k$, $x_{k+1}$, and $y_{k+1}$ have been replaced by $U_0$, $X_0$, $X_1$ and $Y_0$, respectively, and has the solution:
\begin{equation}
\label{eq:iodmd_soln}
\begin{bmatrix} A & B \\ C & D \end{bmatrix} = \begin{bmatrix} X_1 \\ Y_0 \end{bmatrix} \begin{bmatrix} X_0 \\ U_0 \end{bmatrix}^+.
\end{equation}
This procedure is equivalent to the Direct N4SID algorithm \cite[Ch.~6.6]{VanD93,Kat05} mentioned in \cref{sec:intro}.


Note that all the DMD variants discussed so far identify the original continuous-time systems by \emph{linear} discrete-time models.
However, one can create a continuous-time model given by $\{\widehat A, \widehat B, \widehat C, \widehat D\}$, that is a first-order approximation to the discrete-time model obtained by ioDMD, using for example,  
the standard first-order Euler method:
\begin{align*}
 x_{k+1} &= x_k + h (\widehat A x_k + \widehat B u_k) \\
 \Rightarrow A x_k + B u_k &= x_k + h \widehat A x_k + h \widehat B u_k \\
 \Rightarrow \begin{bmatrix} \widehat A & \widehat B \end{bmatrix} &= \begin{bmatrix} h (I - A) & h B \end{bmatrix}.
\end{align*}
The output and feed-through operators for the continuous-time model remain the same as in the discrete-time model produced by ioDMD, that is, $\widehat C = C$, $\widehat D = D$.
Finally, it is important to note that (io)DMD derived models are generally only valid for the time horizon over which the data has been gathered.


\subsection{Reduced Order DMD}\label{sec:rom}
To accelerate the computation of ioDMD-derived models, we follow \cite{morBruJOetal16,morAnnS17} and reduce the order of the possibly high-dimensional state trajectories using a projection-based approach.
The data matrices $X_0$ and $X_1$ are compressed using a truncated (Galerkin) projection $Q \in \R^{N \times n}$, $n \ll N$, $Q^* Q = I$:
\begin{align*}
 \begin{bmatrix} Q^* X_1 \\ Y_0 \end{bmatrix} &= \begin{bmatrix} A_r & B_r \\ C_r & D_r \end{bmatrix} \begin{bmatrix} Q^* X_0 \\ U_0 \end{bmatrix} \\
 \begin{bmatrix} Q^* X_1 \\ Y_0 \end{bmatrix} \begin{bmatrix} Q^* X_0 \\ U_0 \end{bmatrix}^+ &= \begin{bmatrix} A_r & B_r \\ C_r & D_r \end{bmatrix}.
\end{align*}
The order of the identified system is thus determined by the rank of the projection.

A popular method to compute such a truncating projection $Q$ is proper orthogonal decomposition (POD) \cite{morHolLBetal12},
which is practically obtained as the left singular vectors (POD modes) of the data matrix $X$:
\begin{align*}
 X \stackrel{\text{SVD}}{=} U D V^* \rightarrow X_r = U_1^\intercal X \approx X.
\end{align*}
The number of resulting computed POD modes $n$ depends on the desired projection error $\|X - U U^* X\|\fro \leq 
(\sum_{i=1}^n D_{ii}^2)^{1/2}$,
which is a consequence of the Schmidt-Eckhard-Young-Mirsky Lemma (see for example \cite{morAnt05}).

Note again that this data compression scheme has a very different motivation compared to that of the aforementioned dimensionality-reduction done when computing the pseudoinverse via the truncated SVD (discussed in \cref{sec:alg1}).
The latter truncation, based on the magnitude of the singular values, is done for reasons of numerical accuracy when computing the pseudoinverse (and applying it in subsequent computations).
The projection-based truncation discussed here, using the sum of the singular values squared, allows for the possibility of a much less onerous computational burden, as the state space of the models can often be greatly reduced by discarding all but a handful of essential modes in order to obtain a desired approximation error.   
Other projection-based data-driven model reduction techniques for the compression of the state trajectory are similarly applicable, for example empirical balanced truncation \cite{morLalMG99}.

\section{Training Data and Generic Identification}\label{sec:dat}
DMD is a data-driven method, hence, the source of the data used for the system identification needs to be considered.
Usually it is possible to identify an input-output system (for provided discrete input, state, and output functions)
so that the identified system produces similar outputs given the input used for the identification.
To identify a model from data, the associated system needs to produce outputs approximating the data source, not only for specific data sets, but for a whole class of relevant input functions and perhaps even arbitrarily admissible ones.
For such generic linear system identification, 
the matrices $A$, $B$, $C$, $D$ have to be computed in such a manner that
(a) does not require a specific data set and 
(b) allows behaviors of the system being modeled to be predicted for a given initial condition and input function.
This can be accomplished, for example, by persistent excitation (PE) \cite{AstE71,morProBK16}, which utilizes signals such as step functions or Gaussian noise as training input data.
Here, we propose a related approach that also exploits random (Gaussian) sampling,
yet is based on the cross operator.

\subsection{Cross Excitation}\label{sec:wx}
The cross operator \cite{morIonFS09} $W_X : \mathbb{R}^N \to \mathbb{R}^N$ is a tool for balancing-type model reduction and encodes the input-output coherence of an associated system,
which for linear time-invariant systems is the so-called cross Gramian matrix \cite{morFerN83}.
This operator is defined as the composition of the controllability operator $\mathcal{C}: L_2 \to \R^N$ with the observability operator $\mathcal{O}:\R^N \to L_2$:
\begin{align*}
 W_X = \mathcal{C} \circ \mathcal{O}.
\end{align*}
Thus, for a square system with the same input and output space dimension, $W_X$ maps a given initial state $x_0$ via the observability operator to an output function,
and is in turn passed to the controllability operator as an input function resulting in a final state\footnote{This is also illustrated in \cite[Fig.~1]{morIonFS09}}:
\begin{align*}
 x_0 \stackrel{\mathcal{O}}{\mapsto} y \stackrel{\mathcal{C}}{\mapsto} x(T).
\end{align*}
To generate trajectory data, we modify the cross operator by replacing the controllability operator with the input-to-state map $\xi : L_2 \to L_2$  \cite[Ch.~4.3]{morAnt05}.
This yields an operator $\mathcal{W}_X : \mathbb{R}^N \to L_2$:
\begin{align*}
 x_0 \stackrel{\mathcal{O}}{\mapsto} y \stackrel{\xi}{\mapsto} x,
\end{align*}
which maps, as before, an initial state to an output function, but then maps this output (as an input) function to a state trajectory (instead of to a final state).
Compared to PE, the cross(-operator-based) excitation (CE) is a two-stage procedure using perturbations of the initial state to generate the excitation, as opposed to perturbing the input directly.

The cross excitation is related to indirect identification of closed-loop systems \cite{VanS93,OkuF04},
which is also a two-stage process.
First, an intermediary system (open-loop) system is identified,
which then is used in the second step to generate a signal that acts as an excitation for the identification of the actual closed-loop system.
A distinct difference of CE compared to the indirect closed-loop identification approach is 
that the latter exclusively uses input-output data while the former also uses state trajectory data in addition to the input-output data.

\section{Stabilization}\label{sec:stab}
As DMD and its variants are time-domain methods (ioDMD included),
it is typically desired to preserve stability in the (reduced) identified discrete-time systems.
However, models derived by ioDMD are not guaranteed to be stable.
To enforce stability, an additional post-processing step is required.
For example, \cite{morAmsF12} proposed stabilizing models derived using Petrov-Galerkin projections by solving a sequence of semidefinite programs.  In this paper, we take a much more direct approach.

A square matrix $A$ is discrete-time stable if its spectral radius is less than one, that is, $\rho(A) < 1$, where
\[
	\rho(A) \coloneqq \max \{ | \lambda | : \lambda \in \Lambda(A)\}.
\]
Although the spectral radius is nonconvex, it is a continuous function with respect to the matrix $A$ and furthermore, 
it is continuously differentiable \emph{almost everywhere} (in the mathematical sense).
In other words, the set of points where the spectral radius is nonsmooth only has measure 0
and so it holds that points chosen randomly will, with probability 1, be outside of this set.
Hence, despite the nonsmoothness of the spectral radius,
it is still possible to attain a wealth of information from its gradient, 
since it is defined on all but a subset of measure 0 in the full space.
Thus if the matrix $A$ from the ioDMD-derived model, that is, from \eqref{eq:iodmd_soln}, is not stable, 
we could consider employing a gradient-based optimization method to stabilize it, 
while hopefully ensuring that the resulting stabilized version of the ioDMD solution still models the original large-scale system.
In order to meet these two objectives, we consider solving the following constrained optimization problem:
\begin{align}
	\label{eq:conopt}
 	\begin{bmatrix} A_\mathrm{s} & B_\mathrm{s} \\ C_\mathrm{s} & D_\mathrm{s} \end{bmatrix} 
 	= \argmin_{\tilde A, \tilde B, \tilde C, \tilde D}
	\bigg\|\begin{bmatrix} X_1 \\ Y_0 \end{bmatrix} - \begin{bmatrix} \tilde{A} & \tilde{B} \\ \tilde{C} & \tilde{D} \end{bmatrix} 		\begin{bmatrix} X_0 \\ U_0 \end{bmatrix}\bigg\|\fro^2 
	\quad \text{s.t.} \quad 
	|\lambda(\tilde{A})| < 1-\tau,
\end{align}
where $\tilde A \in \R^{r \times r}$, $\tilde B \in \R^{r \times m}$, $\tilde C \in \R^{p \times r}$, $\tilde D \in \R^{p \times m}$, and $\tau \ge 0$ is a margin for the stability tolerance.  
As the unstabilized ioDMD model is already a solution to \eqref{eq:iodmd_eqns}, 
solving \eqref{eq:conopt} should promote solutions that are still close to the original ioDMD model
while simultaneously enforcing the requirement that these models must now be stable, 
due to the presence of the stability radius in the inequality constraint.  
Furthermore, the unstabilized ioDMD model should make a good point to start the optimization method at.

There are however some difficulties with solving \eqref{eq:conopt} iteratively via optimization techniques.
The first is that the objective function of \eqref{eq:conopt} is typically underdetermined in DMD settings, 
which can adversely impact a method's usual rate of convergence, as minimizers are no longer locally unique.
However, as our goal is mainly to stabilize the ioDMD model, without changing its other properties too much, 
we do not \emph{necessarily} need to solve \eqref{eq:conopt} exactly.  
A few iterations may be all that is needed to accomplish this task.

As an alternative, one could consider solving 
\begin{align}
	\nonumber
 	\min_{\tilde A, \tilde B, \tilde C, \tilde D}
	\bigg\|\begin{bmatrix} A_\mathrm{DMD} & B_\mathrm{DMD} \\ C_\mathrm{DMD} & D_\mathrm{DMD} \end{bmatrix} - \begin{bmatrix} \tilde{A} & \tilde{B} \\ \tilde{C} & \tilde{D} \end{bmatrix} 	
	\bigg\|\fro^2 	
	\quad \text{s.t.} \quad 
	|\lambda(\tilde{A})| < 1-\tau
\end{align}
in lieu of \eqref{eq:conopt}, 
where $A_\mathrm{DMD}$, $B_\mathrm{DMD}$, $C_\mathrm{DMD}$, and $D_\mathrm{DMD}$ are the matrices produced by ioDMD.
On the upside, this helps to avoid the problem of underdeterminedness arising in \eqref{eq:conopt} and 
encourages that a stable solution close to the original ioDMD-derived system is found.
However, this modified objective no longer takes any observed data into account.
We did evaluate this alternate optimization problem in our experiments,
but the performance of the models it produced was sometimes worse.
As such, we will only report results for our experiments done using \eqref{eq:conopt} in \cref{sec:numex}.

The second issue for trying to solve \eqref{eq:conopt}
is that the spectral radius can be a rather difficult function to optimize, relatively speaking.
First, despite being continuously differentiable almost everywhere, 
the spectral radius is still a nonsmooth function, specifically at matrices which have multiple eigenvalues that attain the maximum modulus, 
that is, the value of the spectral radius.  
Typically, minimizers of the spectral radius will be at such matrices 
(for example, see the plots of spectral configurations post optimization in 
\cite[Section~4.1 and Appendix~9.1]{CurMO17}),
so optimizing the spectral radius often means that an optimization method must try to converge to a nonsmooth minimizer,
a difficult prospect.
Worse still is that the spectral radius is also non-locally Lipschitz at matrices 
where these multiple eigenvalues attaining the value of the spectral radius coincide (see \cite{BurO01}).  
Many of the available continuous optimization methods are designed under the assumption that functions they optimize are smooth or if not, at least locally Lipschitz.  
If the functions being optimized do not meet these criteria, these methods typically break down.
Furthermore, the nonconvexity of the spectral radius means that optimization codes 
may get stuck in local minima that may or may not provide sufficiently acceptable solutions.

Although the inclusion of the spectral radius constraint makes \eqref{eq:conopt} a nonsmooth, nonconvex optimization problem,
with a non-locally-Lipschitz constraint function, again we do not necessarily need to solve it exactly 
(though this will remain to be seen in our experiments).
Furthermore, while much of the literature on nonsmooth optimization has historically focused on unconstrained problems,
there has also been recent interest in addressing problems with nonsmooth constraints.  
For example, a new algorithm combining quasi-Newton BFGS (Broyden-Fletcher-Goldfarb-Shanno) updating  
and SQP (sequential quadratic programming)\footnote{
For a good introductory reference on many of the optimization techniques referred to in this paper, 
see \cite{NocW99}.}
was recently proposed for 
general nonsmooth, nonconvex, constrained optimization problems \cite{CurMO17},
where no special knowledge or structure is assumed about the objective or constraint functions.
Particularly relevant to our approach here, this BFGS-SQP method was evaluated on 100 different spectral radius constrained optimization problems, with promising results relative to other solvers \cite[Section 6]{CurMO17}.  
This indicates that it may also be a good solver for our nonsmooth constrained optimization problem
and thus we propose using it to solve \eqref{eq:conopt}.
Specifically, we use GRANSO: GRadient-based Algorithm Non-Smooth Optimization
\cite{granso}, an open-source MATLAB code implementing the aforementioned BFGS-SQP method of \cite{CurMO17}.

\section{Numerical Results}\label{sec:numex}

We implemented our new ioDMD variant using both PE and CE sampling strategies 
(presented in \cref{sec:wx}) to collect observations of the original system's behaviors.  
Furthermore, our software can also optionally stabilize the resulting ioDMD-derived models by using GRANSO \cite{CurMO17} 
on our associated nonsmooth constrained optimization problem.

As discussed in \cref{sec:stab}, 
\eqref{eq:conopt} is an underdetermined optimization problem in DMD settings.
Since such problems are in a sense quite flat, 
the norm of the gradient merely being small can be a poor measure of when to terminate;
correspondingly, we tightened GRANSO's default termination tolerance of $10^{-6}$ to $10^{-8}$ (i.e. \texttt{opts.opt\_tol = 1e-8}).
Relatedly, convergence can also be slow so the choice of a starting point can also be critical.
As our goal, specified by \eqref{eq:conopt}, is to stabilize a model 
while minimizing the tradeoff in any increased approximation error (that may occur due to stabilization),
we simply used the unstable ioDMD-derived models as starting points for GRANSO 
and used GRANSO's custom termination feature to halt optimization once a model
had been found that was both stable (for \eqref{eq:conopt} we used $\tau \coloneqq 0$) 
and had an objective value that was less than 1000 times the objective function evaluated at the original ioDMD-derived unstable model.
We found that this loose limit on how much the objective function was allowed to increase
was more than adequate to retain good output errors.
We ran GRANSO using its full-memory BFGS mode (its default behavior and the recommended choice for nonsmooth problems)
and kept all other GRANSO options at their default values as well.

The numerical experiments were implemented in the Matlab language 
and were run under MATLAB R2017a on a workstation computer with an Intel Core i7-6700 (4 Cores @ 3.4 GHz) 
and 8 GB memory.


\subsection{Excitation and Stabilization Evaluation}
\label{sec:numex_eval}
We demonstrate the effects of different types of excitation used for the ioDMD-based system identification by a numerical example.
Specifically, for a given target data set, we identify a discrete-time linear system first using the target data itself,
second by persistent excitation (PE) and third by utilizing the herein proposed cross excitation (CE) from \cref{sec:wx}.

The considered input-output system is based on the transport equation,
with the left boundary of the domain selected as input and the right boundary as output:
\begin{align*}
 \frac{\partial}{\partial t}z(x,t) &= a \frac{\partial}{\partial x}z(x,t), \quad x \in [0,1], \\
 z(0,t) &= u(t), \\
 z(x,0) &= 0, \\
 y(t) &= z(1,t).
\end{align*}
The partial differential equation is discretized in space using the first-order finite-difference upwind scheme,
with a spatial resolution of $\Delta x = \frac{1}{1000}$ and $a = 1.3$.
The resulting ODE input-output system is then given by:
\begin{align}\label{eq:lintrans}
\begin{split}
 \dot{x}(t) &= Ax(t) + Bu(t), \\
       y(t) &= Cx(t).
\end{split}
\end{align}
The target data is given by discrete input, state and output functions from a simulation, which is performed by a first order implicit Runge-Kutta method.  We used a time-step width of $\Delta t = \frac{1}{1000}$ and a time horizon of $T = 1$, with a Gaussian-bell type input function \mbox{$\hat{u}(t) = e^{-\frac{1}{1000}(t-\frac{1}{10})^2}$}.

For the comparison, a discrete-time linear system was first identified from the $1000$ snapshots generated by a simulation of \eqref{eq:lintrans} excited by input $\hat{u}$, to obtain a baseline for the modeling performance of ioDMD.
The PE variant was sampled using zero-mean, unit-covariance Gaussian noise and a unit step function input was used for the ioDMD-based identification.
Finally, our CE variant had the initial state sampled from a unit Gaussian distribution and a (component-wise) shifted initial state $x_{0,i} = 1$ was tested.
All methods were tested with ioDMD only and then also with stabilization.

\subsubsection{ioDMD without Stabilization}\label{sec:numex1}
\begin{figure}[h!]
\begin{subfigure}[t]{0.48\textwidth}\centering
 \includegraphics[width=\textwidth]{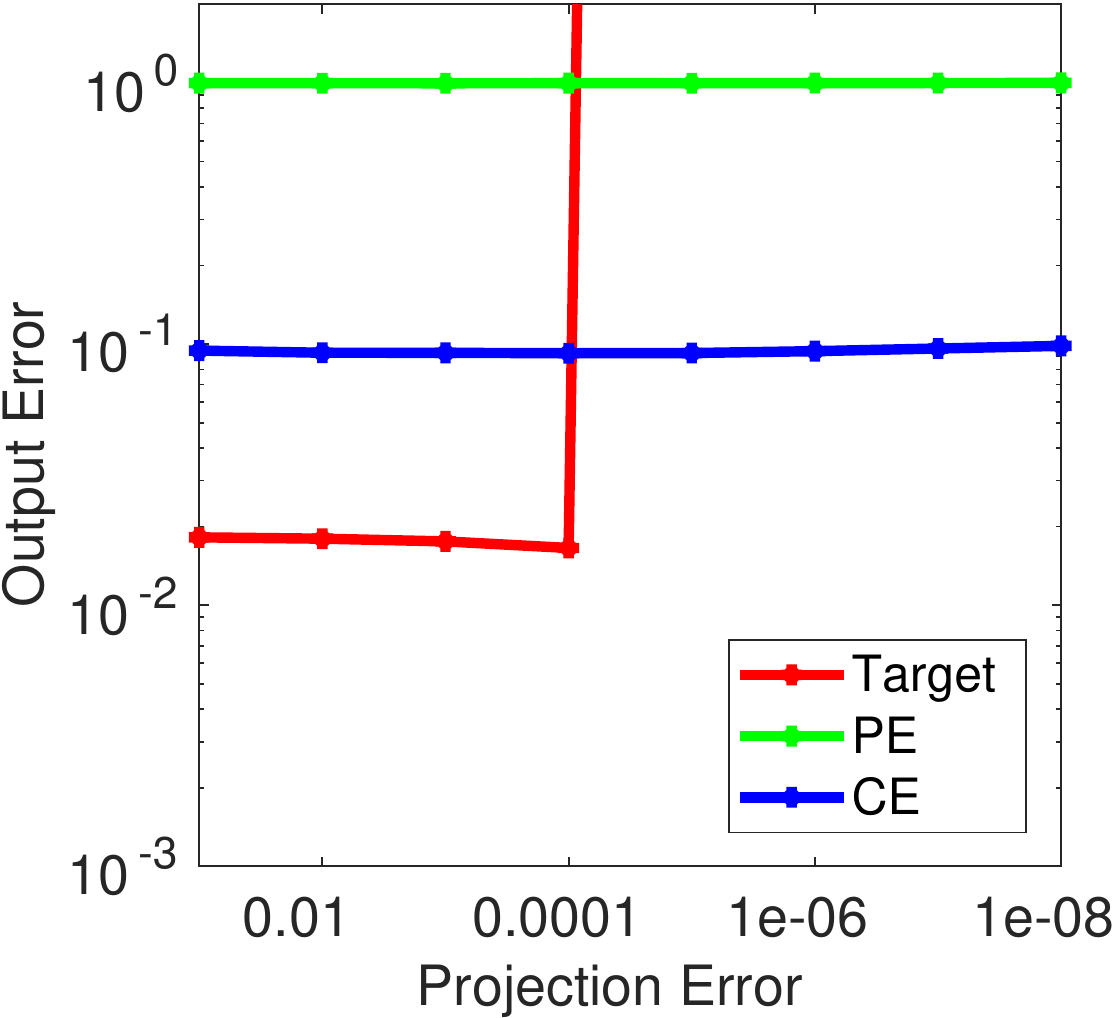}
 \caption{Non-regularized identification by noise input or randomly sampled initial state.}
 \label{fig:numex1a}
\end{subfigure}
~
\begin{subfigure}[t]{0.48\textwidth}\centering
 \includegraphics[width=\textwidth]{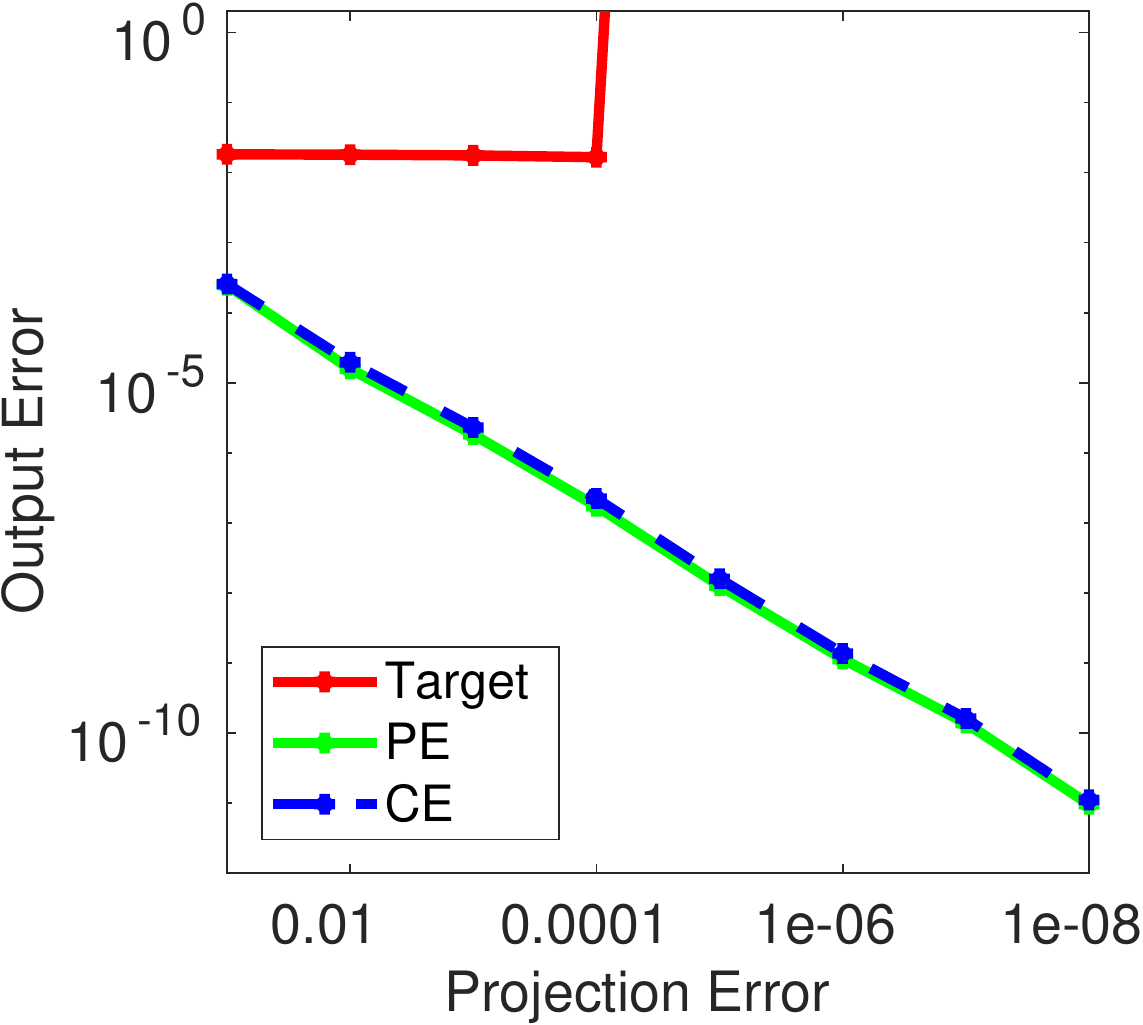}
 \caption{Non-regularized identification by step input or shifted initial state.}
 \label{fig:numex1b}
\end{subfigure}

\begin{subfigure}[t]{0.48\textwidth}\centering
 \includegraphics[width=\textwidth]{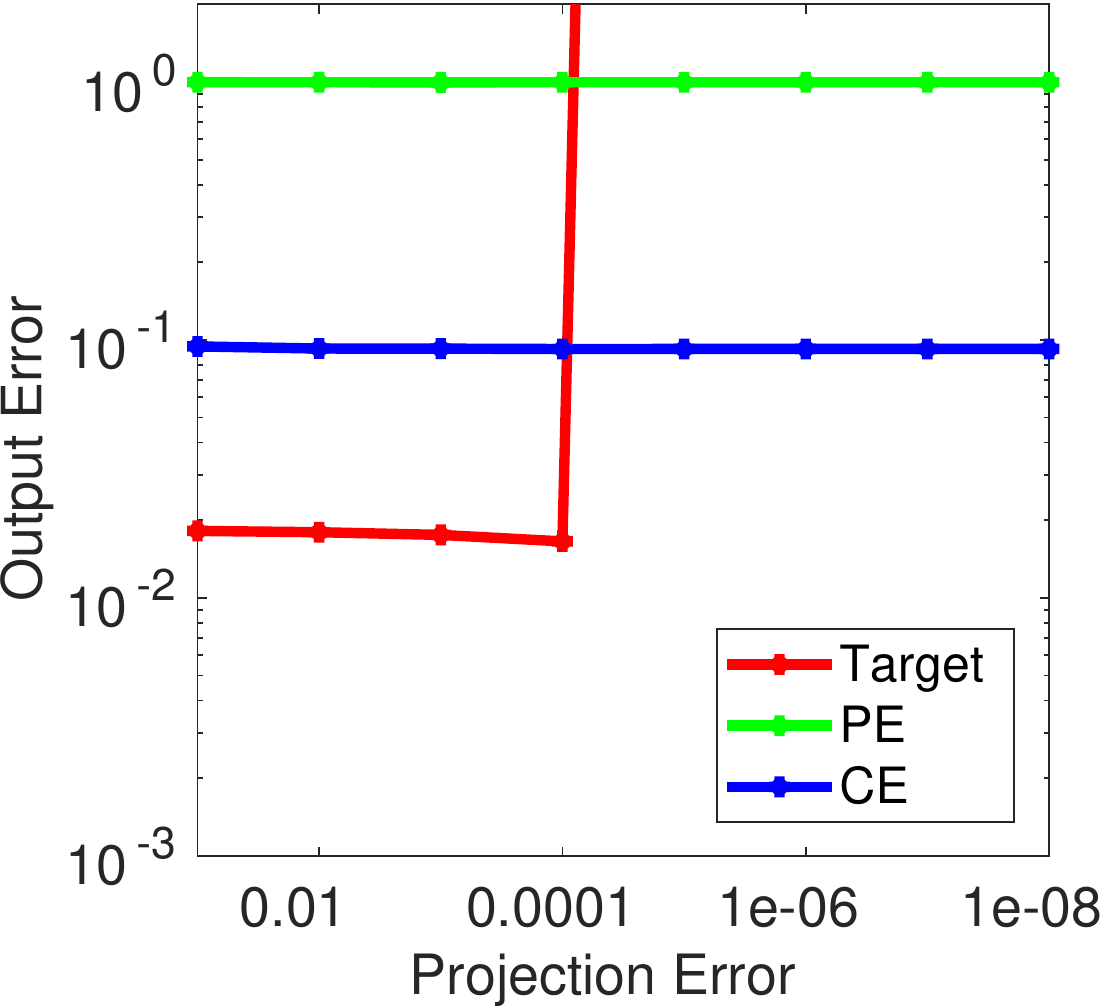}
 \caption{Regularized identification by noise input or randomly sampled initial state.}
 \label{fig:numex1c}
\end{subfigure}
~
\begin{subfigure}[t]{0.48\textwidth}\centering
 \includegraphics[width=\textwidth]{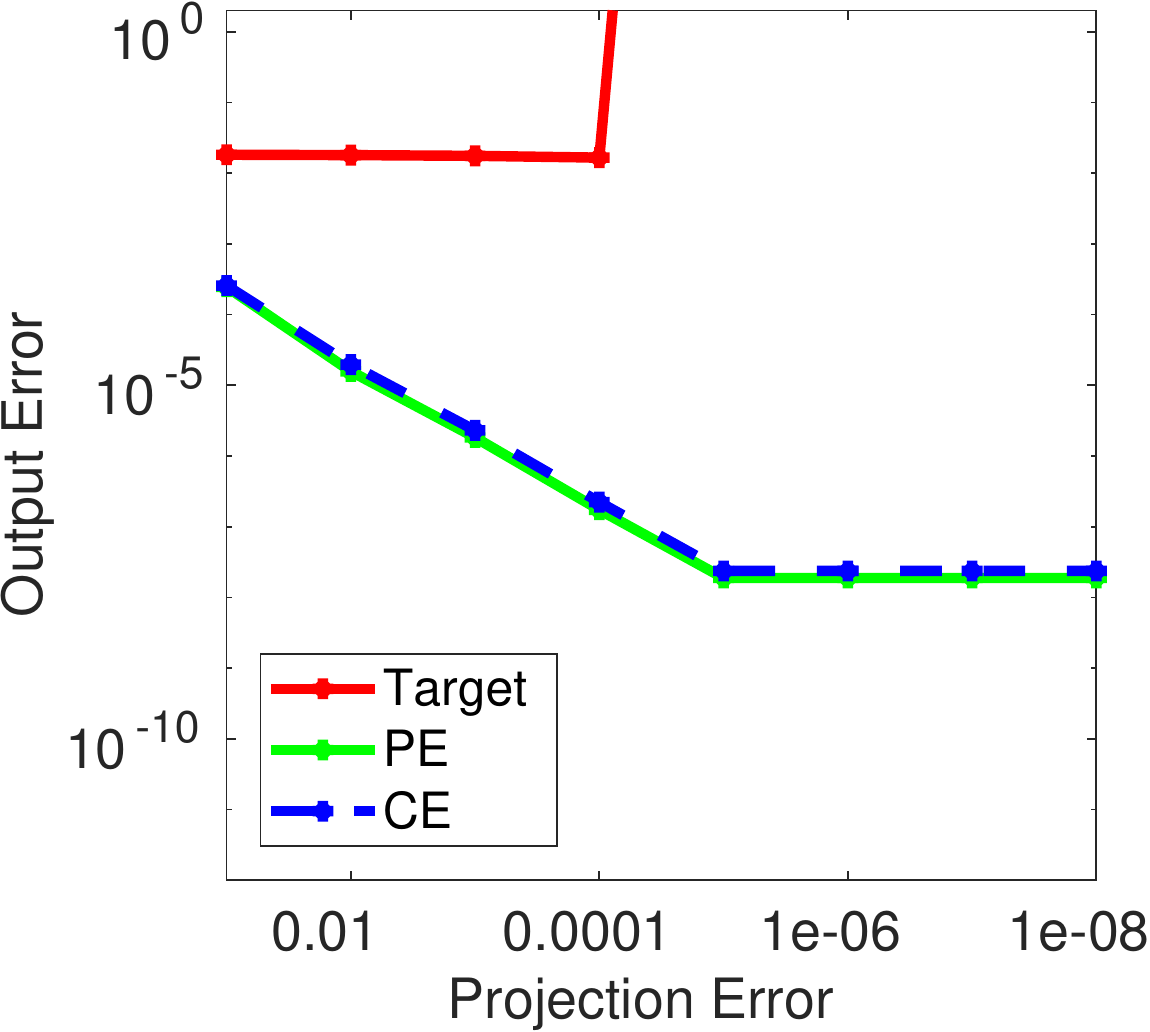}
 \caption{Regularized identification by step input or shifted initial state.}
 \label{fig:numex1d}
\end{subfigure}

\caption{First numerical experiment (\cref{sec:numex1}): Non-stabilized ioDMD-based system identification. The plots show the identified (reduced order) system's output error compared to the original system's output for varying accuracies of the POD projection-based model reduction.}
\label{fig:numex1}
\end{figure}

\cref{fig:numex1} depicts the relative output error for simulations of systems identified using data associated to the target input $\hat{u}$, PE, and CE for increasingly accurate state-space data compression (that is, for increasingly smaller amounts of compression).
The state-space data dimensionality was reduced using the POD method with prescribed projection errors of $10^{-1}, \dots, 10^{-8}$.
In this set of experiments, we did not use stabilization.

In \cref{fig:numex1a,fig:numex1b}, the ioDMD procedure was not regularized by truncating singular values,
while regularization $\Sigma_{ii} < 10^{-5}$ was used in \cref{fig:numex1c,fig:numex1d}.  
In \cref{fig:numex1a,fig:numex1c}, system identification was performed using zero-mean Gaussian noise for the PE and an initial state sampled from a zero-mean Gaussian distribution for CE, respectively.  
In \cref{fig:numex1b,fig:numex1d}, respectively, the identification was driven by a step input for the PE and a shifted initial state $x_{0,i} = 1$ for CE.

For this set of experiments, using the target data only produced stable systems for large values of acceptable projection error
while the PE-derived models were always unstable (and thus had very poor performance regardless of the projection error).
In contrast, the CE method produced stable systems for all levels of projection error tested.
Performance-wise, when comparing to the few target-data-derived models that also happened to be stable, 
the CE-derived models had errors that were less than one order of magnitude higher (see \cref{fig:numex1a,fig:numex1c}).
On the other hand, when using the step input or shifted initial state, both PE and CE produced models with increasing accuracy as the level of  acceptable projection error of the data was decreased, as seen in \cref{fig:numex1b,fig:numex1d}.
In \cref{fig:numex1d}, we see that the regularization limited the attainable accuracy for both PE and CE.
The target-data-derived system had a constant error independent from the projection error of the data.

\subsubsection{ioDMD with Stabilization}\label{sec:numex2}
\begin{figure}[h!]
\begin{subfigure}[t]{0.48\textwidth}\centering
 \includegraphics[width=\textwidth]{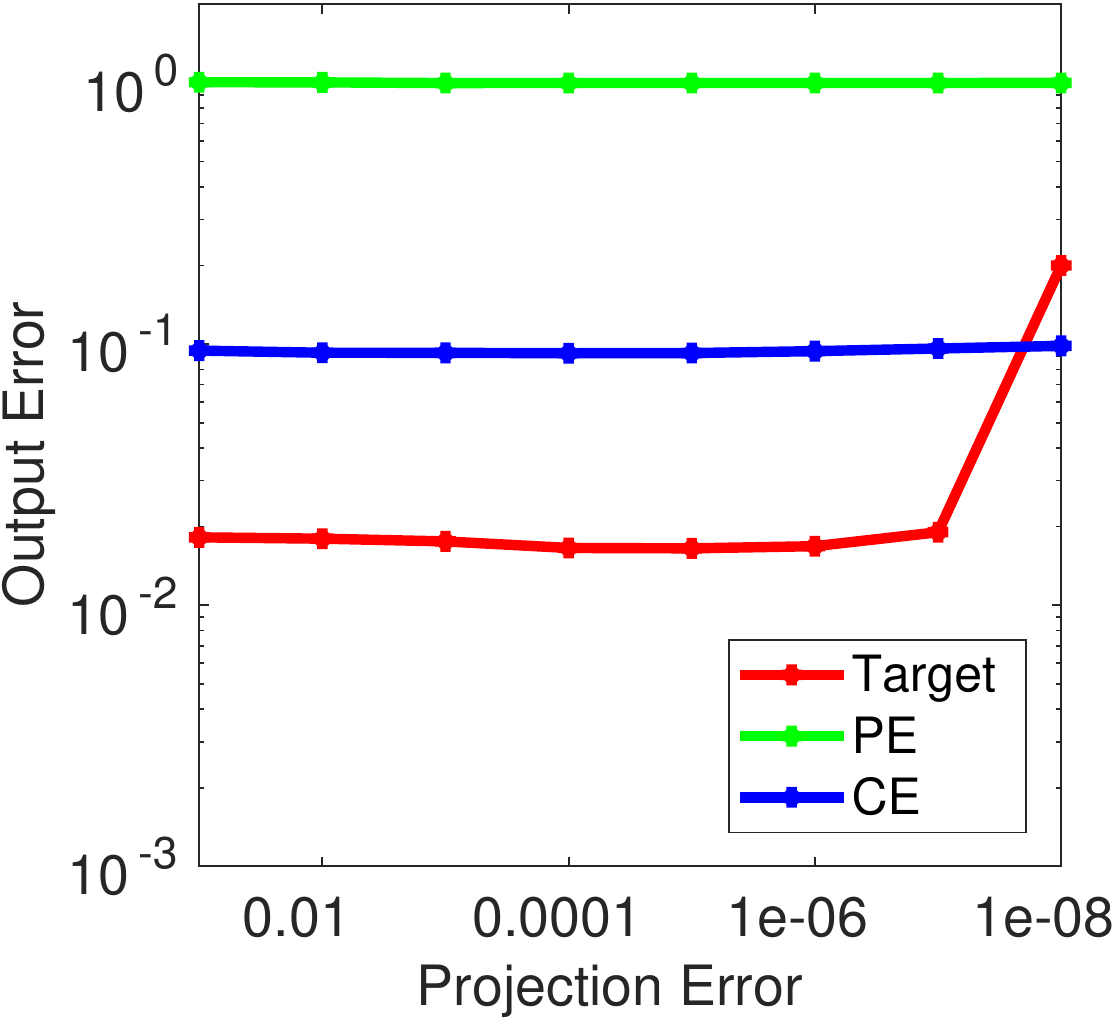}
 \caption{Non-regularized identification by noise input or randomly sampled initial state.}
 \label{fig:numex2a}
\end{subfigure}
~
\begin{subfigure}[t]{0.48\textwidth}\centering
 \includegraphics[width=\textwidth]{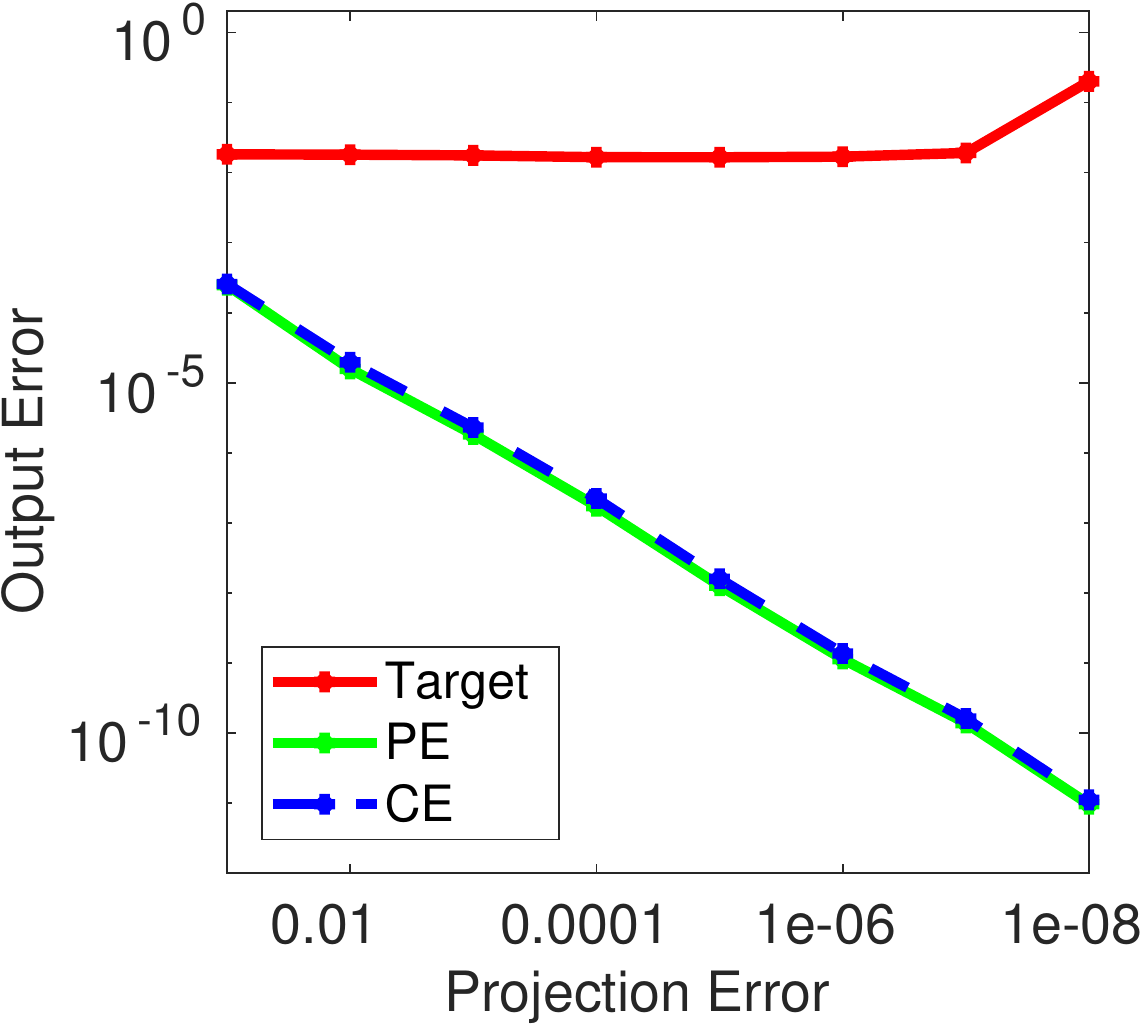}
 \caption{Non-regularized identification by step input or shifted initial state.}
 \label{fig:numex2b}
\end{subfigure}

\begin{subfigure}[t]{0.48\textwidth}\centering
 \includegraphics[width=\textwidth]{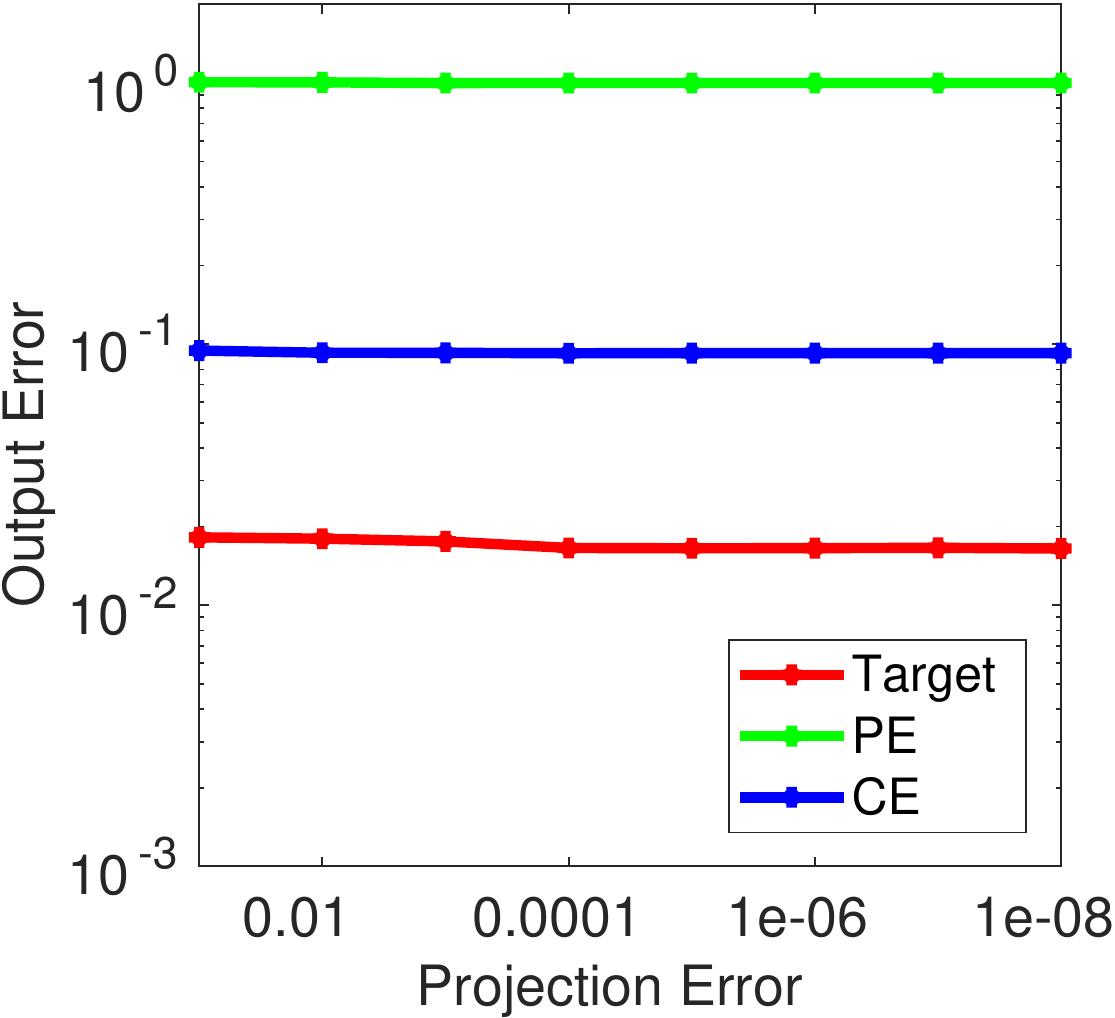}
 \caption{Regularized identification by noise input or randomly sampled initial state.}
 \label{fig:numex2c}
\end{subfigure}
~
\begin{subfigure}[t]{0.48\textwidth}\centering
 \includegraphics[width=\textwidth]{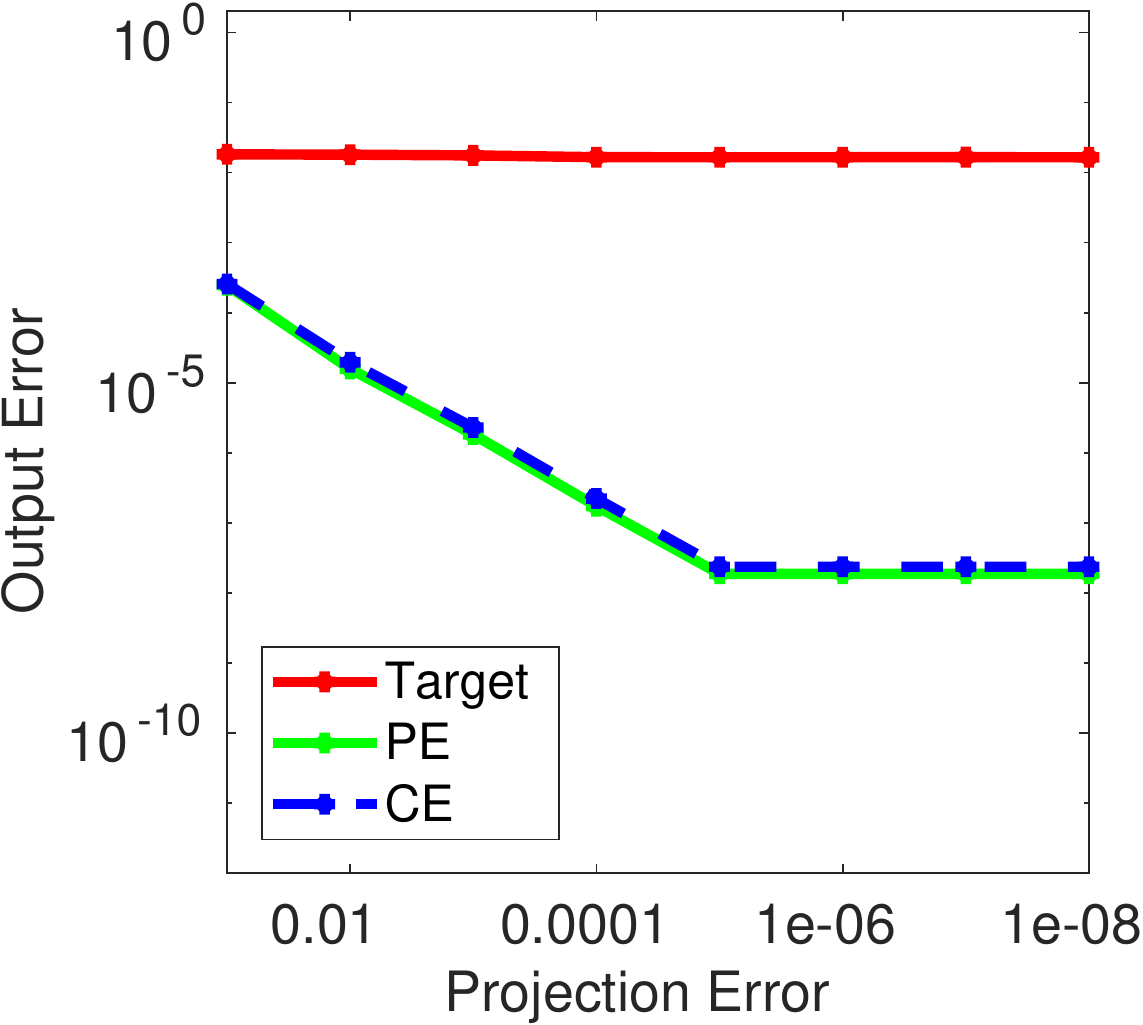}
 \caption{Regularized identification by step input or shifted initial state.}
 \label{fig:numex2d}
\end{subfigure}

\caption{Second numerical experiment (\cref{sec:numex2}): Stabilized ioDMD-based system identification. The plots show the identified (reduced order) system's output error compared to the original system's output for varying accuracies of the POD projection-based model reduction.}
\label{fig:numex2}
\end{figure}

In this second set of experiments,
\cref{fig:numex2} still shows the relative output error for simulations of systems identified using the target data, PE, and CE for increasingly accurate state-space data compression, but now the systems have been post-processed using our optimization-based approach to enforce stability, as necessary.
The state-space data dimensionality was again reduced using the POD method, with prescribed projection errors of $10^{-1}, \dots, 10^{-8}$.  The subfigures are arranged as they were in \cref{fig:numex1}.

The step function PE and shifted initial state CE are unaffected by our stabilization post-processing phase,
as these systems were already stable; thus their plots are the same in \cref{fig:numex2b,fig:numex2d} as they were in  \cref{fig:numex1b,fig:numex1d}.
In the case of using Gaussian noise or randomly sampled initial state (\cref{fig:numex2a,fig:numex2c}), which had not yielded stable systems for the target data or PE (either with or without regularization), 
our optimization-based post-processing procedure now enforced stability.  

For the ioDMD-derived models that were unstable, GRANSO was able to stabilize all 24 of them.
The average number of iterations needed to find the first stabilized version was 13.5 while the maximum was just 61.
Furthermore, for 12 of the problems, our full termination criteria were met in less than 20 iterations
while the average and maximum iteration counts over all 24 problems were respectively 84.2 and 329.
This demonstrates that our optimization-based approach is indeed able to stabilize such models reliably and efficiently.
Solving \eqref{eq:conopt} via GRANSO also met our secondary goal, 
that stabilization is achieved without deviating too significantly from the original unstable models.
The largest observed relative change
between an initial unstable model and its corresponding GRANSO-derived stabilized version was just 1.44\%
while the average observed relative change was merely 0.231\%; 
the relative differences were calculated by 
comparing $\operatorname{vec}[A_\mathrm{s}, B_\mathrm{s}; C_\mathrm{s}, D_\mathrm{s}]$,
where the matrices are GRANSO's computed stabilized solution to \eqref{eq:conopt},
to a similar $\operatorname{vec}$ of the original (unstable) ioDMD-derived model.

\subsubsection{Reduced Orders and Runtimes}
\begin{figure}[h!]
\begin{subfigure}[t]{0.48\textwidth}\centering
 \includegraphics[width=\textwidth]{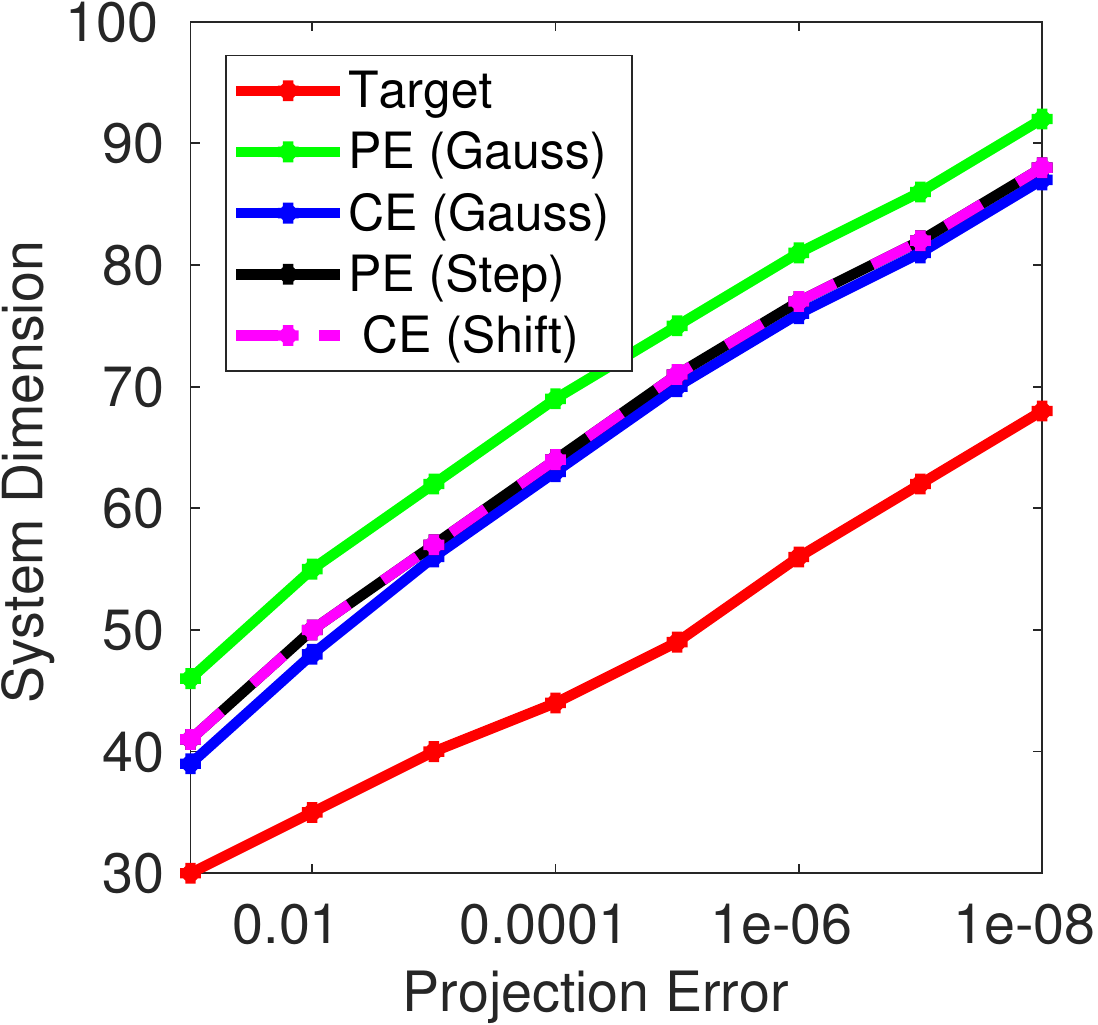}
 \caption{Comparison of the identified system's reduced order for different pre-selected mean $L_2$ projection errors used in both experiments (see \cref{sec:numex1}).}
 \label{fig:numex3a}
\end{subfigure}
~
\begin{subfigure}[t]{0.48\textwidth}\centering
 \includegraphics[width=\textwidth]{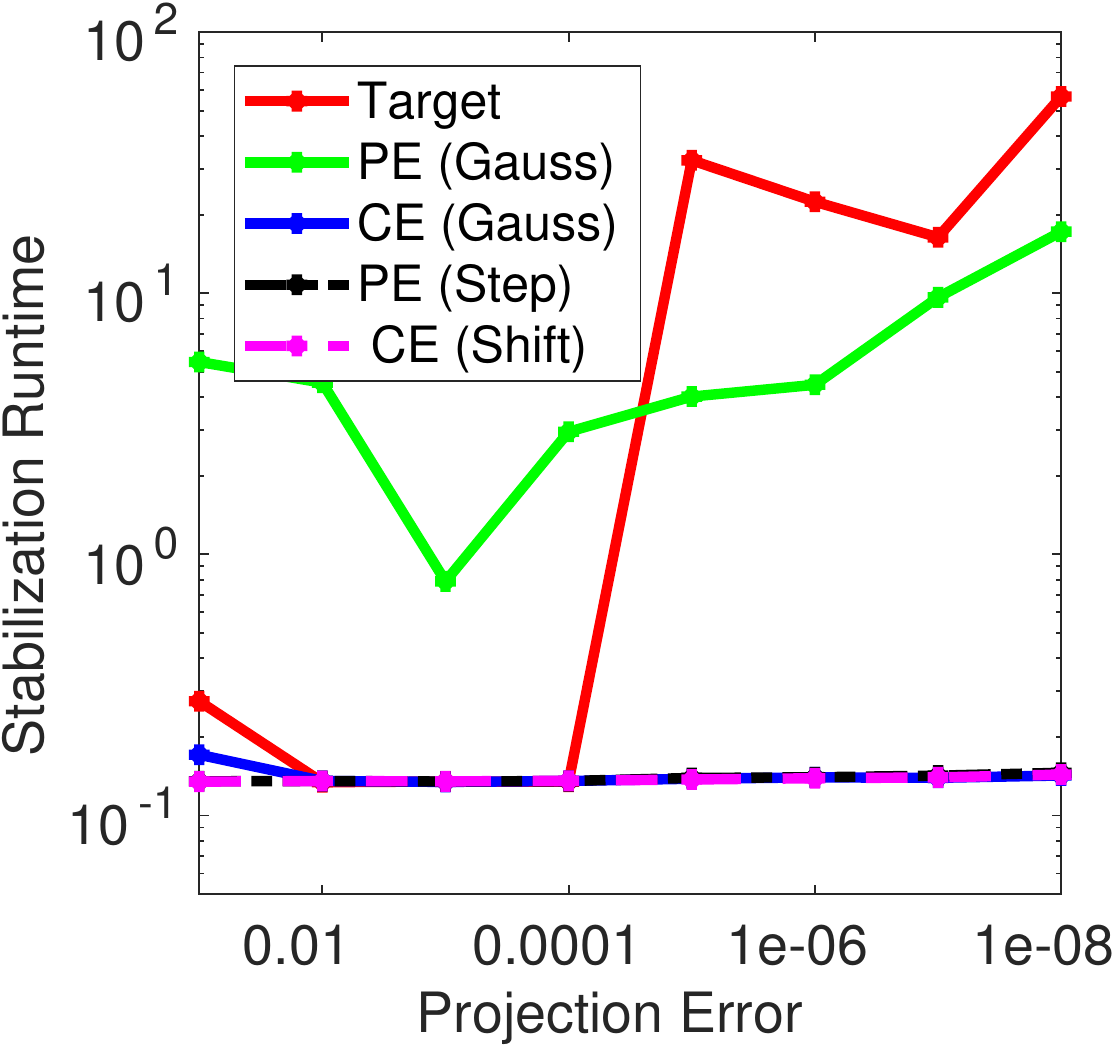}
 \caption{Comparison of runtimes in seconds for the stabilization procedure in the second experiment (\cref{sec:numex2}).}
 \label{fig:numex3b}
\end{subfigure}
\caption{Comparison of reduced identified system orders and stabilization runtimes.}
\label{fig:numex3}
\end{figure}

We now compare the order of the identified systems.
The order of the identified system is determined by the projection error selected for the state-space compression in the POD.
For each data set (Target, Gaussian noise PE, Gaussian sample CE, Step input PE, shifted initial state CE),
the resulting reduced order is plotted for the different prescribed projection errors in \cref{fig:numex3a}.
As we can see, the step-input PE and shifted-initial-state CE method behave similar in terms of system dimension
while for the Gaussian-noise PE and Gaussian-sampled initial-state CE,
the CE variant resulted in smaller system dimensions.

In terms of computational cost, only the Target and Gaussian-noise PE variants required stabilization and as such,
it is only for these that we see increased runtimes, as shown by the red and green plots (respectively) in \cref{fig:numex3b}.
Otherwise, the runtimes were mostly identical for the other variants in the comparison.

\subsection{Limited-Memory BFGS for Stabilization?}
One potential downside to our optimization-based stabilization procedure is that 
\emph{full-memory} BFGS inverse Hessian updating 
(GRANSO's recommended setting for nonsmooth problems)
requires per-iteration work and storage that is quadratic in the number of optimization variables.
As the number of optimization variables is $(r+m)\times(r+p)$, 
the running time required to solve \eqref{eq:conopt} could become unacceptably long 
as $r$, the reduced order of the model, is increased.
Thus, we now also consider whether or not \emph{limited-memory} 
BFGS updating can also be effective for solving \eqref{eq:conopt}.

Using limited-memory BFGS has the benefit that the per-iteration work and storage 
is reduced to a linear amount (again in the number of optimization variables).
However, one of the tradeoffs is that convergence can be quite slow in practice.  
For smooth problems, full-memory BFGS converges superlinearly while limited-memory BFGS only does so linearly;
on nonsmooth problems, linear convergence is the best one can typically expect.
Another potential issue is that 
while there has been much evidence supporting that full-memory BFGS is very reliable for nonsmooth optimization,
the case for using limited-memory BFGS on nonsmooth problems is much less clear;
for a good overview of the literature on this topic, see \cite[Section~1]{LewO13}.

To investigate this question,
we reran the experiments from \cref{sec:numex_eval} a second time 
but where GRANSO's limited-memory BFGS mode was now enabled.
Specifically, we configured GRANSO to approximate the inverse Hessian at each iteration using only the 10 most recently computed gradients,
accomplished by setting \texttt{opts.limited\_mem\_size = 10}.
All other parameters of our experimental setup were kept as they were described earlier.

In this limited-memory configuration, 
GRANSO often required significantly more iterations, as one might expect.
The average and max number of iterations to find the first stable version of a model
were respectively 73.6 and 474, 
about an order of magnitude more iterations than incurred when using full-memory BFGS.
On the other hand, for 19 of the 24 problems, stable models were encountered within the first 20 iterations.
To meet our full termination criteria, 
the average and max number of iterations were respectively 222.0 and 811,
roughly about two and a half times more than incurred when using full-memory BFGS.
Nevertheless, 12 of the 24 problems were still satisfactorily solved in less than 20 iterations,
matching the earlier result when using full-memory BFGS.
Despite the large increases in iteration counts, 
GRANSO's overall runtime was on average 3.83 times faster when enabling limited-memory BFGS.

With respect to output error in this limited-memory evaluation, 
the resulting stabilized models still essentially matched the earlier results using full-memory BFGS.  
There was one notable exception, for Target data using the smallest projection error of $10^{-8}$,
where GRANSO remarkably found a better-performing model when using limited-memory BFGS.
However, we did observe that the quality of the stabilized models appeared to be much more sensitive 
to changing GRANSO's parameters than they were when using full-memory BFGS.
As a consequence, we still advocate that solving \eqref{eq:conopt} with GRANSO is generally best done using 
its default full-memory BFGS updating.
Nonetheless, if this is simply not feasible computationally, 
one may still be able to obtain good results using limited-memory BFGS 
but perhaps not as reliably or consistently.

As a final clarifying remark on this topic, 
we note that one cannot necessarily expect good performance on nonsmooth problems
when using \emph{just any} BFGS-based optimization code
and that generally it is critical that the choice of software is one specifically designed for nonsmooth optimization.
Indeed, this is highlighted in the evaluation done in \cite[Section~6]{CurMO17},
where off-the-shelf quasi-Newton-based codes built for smooth optimization 
perform much worse on a test set of nonsmooth optimization problems
compared to the quasi-Newton-based codes specifically built with nonsmooth optimization in mind.


\section{Conclusion}
In this work, we evaluated the approximation quality of ioDMD system identification using a novel excitation scheme
and a new optimization-based, post-processing procedure to ensure stability of the identified systems.
Our new cross excitation strategy, particularly when used with random sampling, often produces
better results than when using persistent excitation, and our experiments indicate that 
both excitation schemes are useful for efficiently obtaining good models for approximating the target data.   
Furthermore, we show that directly solving a nonsmooth constrained optimization problem can indeed be a viable approach 
for stabilizing ioDMD-derived systems
while retaining the salient properties for approximating the output response.

\vfill

\section*{Code Availability}
The source code of the presented numerical examples can be obtained from:
\begin{center}
\url{http://runmycode.org/companion/view/2902}
\end{center}
and is authored by: \textsc{Christian Himpe} and \textsc{Tim Mitchell}.

\pagebreak


\bibliographystyle{plainurl}      
\bibliography{mor,csc,software}   

%
%

\end{document}